\documentclass[runningheads]{llncs}
\usepackage{amsmath,amsfonts}
\usepackage{verbatim}
\usepackage[T1]{fontenc}
\usepackage{graphicx}
\begin{document}
\title{Passive Radio Frequency-based 3D Indoor Positioning System via Ensemble Learning\thanks{Supported by Air Force Office of Scientific Research.}}
%
%
\author{Liangqi Yuan\inst{1}\orcidID{0000-0002-9994-6773} \and
Houlin Chen\inst{2} \and
Robert Ewing\inst{3} \and
Jia Li\inst{1}}
\authorrunning{L. Yuan et al.}
%
\institute{Oakland University, Rochester, MI 48309, USA 
\email{\{liangqiyuan,li4\}@oakland.edu} \and
University of Toronto, Toronto, ON M5S 1A1, CA 
\email{houlin.chen@mail.utoronto.ca} \and
Air Force Research Laboratory, WPAFB, OH 45433, USA 
\email{robert.ewing.2@us.af.mil}
}
\maketitle              
\begin{abstract}
Passive radio frequency (PRF)-based indoor positioning systems (IPS) have attracted researchers' attention due to their low price, easy and customizable configuration, and non-invasive design. This paper proposes a PRF-based three-dimensional (3D) indoor positioning system (PIPS), which is able to use signals of opportunity (SoOP) for positioning and also capture a scenario signature. PIPS passively monitors SoOPs containing scenario signatures through a single receiver. Moreover, PIPS leverages the Dynamic Data Driven Applications System (DDDAS) framework to devise and customize the sampling frequency, enabling the system to use the most impacted frequency band as the rated frequency band. Various regression methods within three ensemble learning strategies are used to train and predict the receiver position. The PRF spectrum of 60 positions is collected in the experimental scenario, and three criteria are applied to evaluate the performance of PIPS. Experimental results show that the proposed PIPS possesses the advantages of high accuracy, configurability, and robustness.

\keywords{Indoor positioning system \and Passive radio frequency \and Signal of opportunity \and Ensemble learning \and Machine learning.}
\end{abstract}
\section{Introduction}
\label{Sec. Introduction}
Signals of opportunity (SoOP) for implementing indoor positioning system (IPS) has shown progress in recent years \cite{b1,b2}. SoOP refers to some non-task signals that are used to achieve specified tasks, such as Wi-Fi, cellular network, broadcasting, and other communication signals for positioning tasks. These communication signals have different frequencies according to different functions. For example, the frequency of broadcast signals is tens to hundreds of MHz, and the frequency of Wi-Fi can reach 5GHz. Each SoOP has different performances for different tasks, which will be affected by the local base stations, experiment scenarios, and task settings. SoOP aim to facilitate high-precision positioning in GPS-shielded environments while avoiding the need for additional signal sources. However, how to use a single receiver for positioning in an environment where the signal source is unknown is still an open problem. Therefore, a passive radio frequency (PRF) system is proposed to integrate these communication signals due to the design of a customizable frequency band. Finding the frequency band most impacted for positioning is the most significant prior. In addition, PRF can capture scenario signatures, including liquids, metal objects, house structures, etc., which has been proven to further improve the performance of a positioning system.

Dynamic Data Driven Applications System (DDDAS) frameworks have already shown their application prospects, such as in the fields of environmental science, biosensing, autonomous driving, etc. The application of the DDDAS framework to these domains varies, depending on the input variables and output decisions of the system. Table \ref{Table Instantaneous DDDAS vs. Long-Term DDDAS} shows some examples of instantaneous and long-term DDDAS. Currently, most of DDDAS are emphasized to instantaneous DDDAS, which require us to react immediately to dynamic data input. For example, hurricane forecasting is an instantaneous DDDAS, and if it doesn't react in time, there will be some serious consequences. But long-term DDDAS also has its benefits, and there are no serious consequences for not responding immediately, such as an energy analysis DDDAS is used to save consumption. The advantage of long-term DDDAS is dynamic data input, which can effectively reduce consumption, improve accuracy, and enhance robustness.
\begin{table}
\centering
\caption{Instantaneous DDDAS vs. Long-Term DDDAS.}\label{Table Instantaneous DDDAS vs. Long-Term DDDAS}
\begin{tabular}{|c|c|}
\hline
Instantaneous & Long-Term \\
\hline
Weather forecasting \cite{plale2005towards} & Energy analysis \cite{neal2016energy} \\
Atmospheric contaminants \cite{patra2013challenges} & Materials Analysis \cite{mulani2020uncertainty} \\
Wildfires detection \cite{michopoulos2004agent} & Identification of biomarkers in DNA methylation \cite{damgacioglu2022dynamic} \\
Autonomous driving \cite{allaire2013offline} & Multimedia content analysis \cite{blasch2018dynamic} \\
Fly-by-feel aerospace vehicle \cite{kopsaftopoulos2022dynamic} & Image processing \cite{li2018design} \\
Biohealth outbreak \cite{yan2018dynamic} & Our proposed positioning system \\
\hline
\end{tabular}
\end{table}

Due to the uncertainty of SoOPs and scenario signature, IPSs need to conform to the paradigm of DDDAS \cite{b3,b10}. For the PRF positioning system, the selection of frequency band is a dynamic issue, which is determined according to the scenario signature. Therefore, the computational feedback in DDDAS is required to reconfigure the sensor for frequency band selection. Selecting some frequency bands from the full frequency band can effectively save sampling time, computing resources, and increase the robustness, etc. \cite{b4}. Moreover, the customizable frequency band can be used in a variety of different tasks, such as human monitoring, navigation, house structure detection, etc. \cite{b5,mu2021humani,mu2021humanp}. Therefore, the PRF-based systems under the DDDAS framework need to dynamically optimize the frequency parameter according to its usage scenarios and task settings to obtain higher adaptability, accuracy, and robustness.

Ensemble learning is used as the strategy for the positioning regression task due to its ability to integrate the strengths of multiple algorithms \cite{b6,b9}. Ensemble learning includes three strategies, namely boosting, bagging, and stacking \cite{b11}, depending on whether the base estimator is parallel or serial \cite{b7}. The boosting strategy is a serial strategy where the posterior estimator learns the wrong samples of the prior estimator, which reduces the bias of the model. However, this strategy overemphasizes the wrong samples and thus may lead to larger variance and weaker generalization ability of the model. Both bagging and stacking strategies are parallel structures, which can reduce the variance and enhances the generalization ability. Compared to the bagging strategy, which uses averaging as the final estimator, stacking uses a regressor as the final estimator. Compared to linear or weighted averaging, the model can further reduce model bias by analyzing the decisions of the base estimators.

This paper proposes a PRF-based 3D IPS, named PIPS, for the positioning regression task. Within the DDDAS framework, the performance of the PIPS system is enhanced by adaptive frequency band selection, which continues the most impacted frequency band found in the previous work \cite{b4}. PRF spectrum data was collected at 60 gridded positions in the scenario. The spectrum data set for positioning is trained in three ensemble learning strategies. Root mean square error (RMSE) is used to evaluate the accuracy of PIPS, coefficient of determination $R^2$ is used to evaluate the reliability, and 95$\%$ confidence error (CE) is used to evaluate the optimality. Experiments demonstrate that the proposed PIPS exhibits its potential for accurate object locating tasks. 

This paper is presented as follows. In Section II, the details and sensor settings of the proposed PIPS are illustrated. The experimental setup and results are shown in Section III. Section IV gives some discussions on the advantages of PIPS under the DDDAS framework prior to the conclusion and future work demonstrated in Section V.

\section{Frequency-adaptive PIPS}
\label{Sec. Frequency-adaptive PIPS}

\begin{figure}[h]
\centering
\includegraphics[width=1\linewidth]{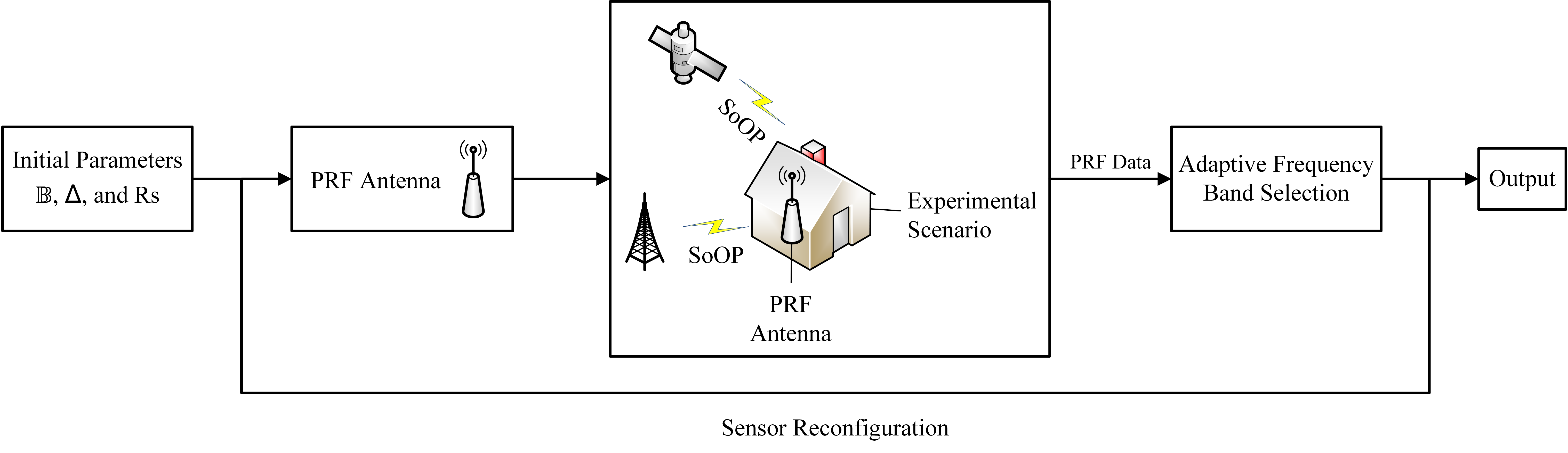}
\caption{DDDAS framework reconfigures the parameters $\mathbb{B}$, $\Delta$, and $R_s$ of the PRF sensor in PIPS.}
\label{Fig. System Diagram}
\end{figure}

PIPS achieves sensing by passively accepting the PRF spectrum in the scenario. Software-defined radio (SDR) is used to control the PRF sensor for data collection, including the frequency band $\mathbb{B}$, step size $\Delta$, sampling rate $R_s$, etc. Reasonable selection of the parameters of the PRF sensor in PIPS is crucial. The diagram of frequency band selection by PIPS under the framework of DDDAS is shown in Fig. \ref{Fig. System Diagram}. The DDDAS framework is used to reconfigure the parameters of the PRF sensor, which is achieved through SDR. The parameters of the PRF sensor, especially the center frequency, are dynamically reconfigured to adapt to the signatures of different scenarios. 

With the support of initial parameters $\mathbb{B}$, $\Delta$, and $R_s$, the data set collected by the PRF sensor $D \in \mathbb{R}^{n \times m}$ and its corresponding position label set $C \in \mathbb{R}^{n \times 3}$. $D$ is the PRF spectrum, that is, the average powers collected over the frequency band. Although it is feasible to use the average power corresponding to the full frequency band as the feature vector for positioning, it will greatly increase the sampling time. Therefore, it is necessary to optimize the initial parameters $\mathbb{B}$, $\Delta$, and $R_s$ under the DDDAS framework. The proposed PIPS system can be defined as the following function $f:C \to D$,
\begin{equation}
d = f(c;\mathbb{B},\Delta,R_s),
\label{Eq. 1}
\end{equation}
where $d$ and $c$ are a pair of samples in $D$ and $C$, which also represent a pair of corresponding PRF spectrum and coordinate. Eq. \ref{Eq. 1} shows the collection of PRF data at the corresponding coordinates given the parameters. By training on the ensemble learning model on the collected data and making predictions, the estimated coordinates can be obtained:
\begin{equation}
\hat{c} = f^{-1}(d;\mathbb{B},\Delta,R_s).
\label{Eq. 1}
\end{equation}

\begin{figure}[!h]
\centering
\includegraphics[width=0.6\linewidth]{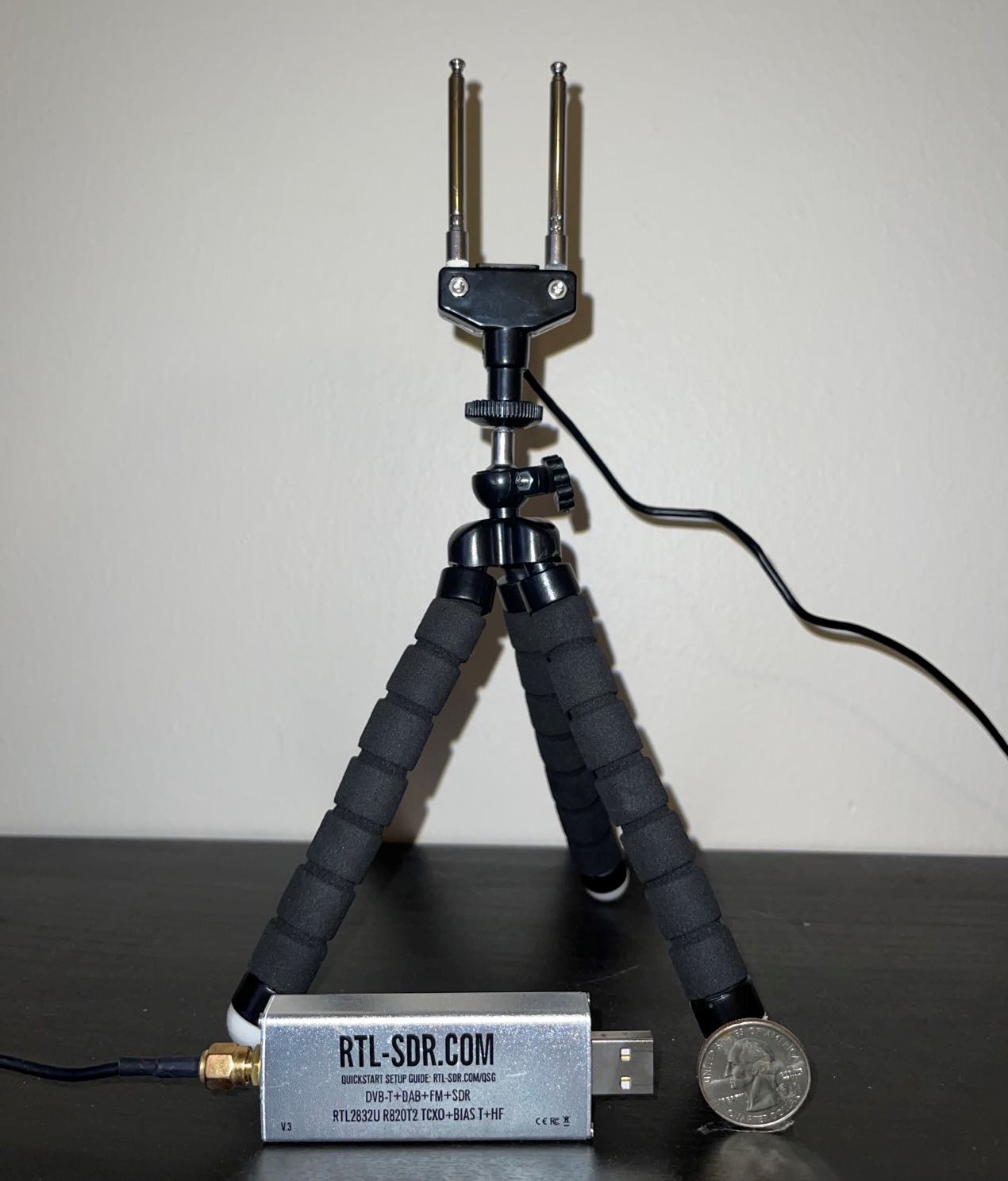}
\caption{RTL-SDR RTL2832U is used as PRF sensor to collect PRF spectrum.}
\label{Fig. PRF Antenna}
\end{figure}

The PRF spectral data collected by the PRF sensor in the experimental scenario contains the SoOP and the signature of the experimental scenario. The PRF sensor in PIPS is reconfigured after the adaptive band selection algorithm is used to find the most impacted band for the positioning task. After the $k$-th optimization, the collected data set under the optimized parameter $\mathbb{B}_k$, $\Delta_k$, and $R_{s_k}$ is defined as $D_k \in \mathbb{R}^{n \times m_k}$. Dynamic reconfiguration may be performed once or multiple times, depending on the properties of the SoOP in the scenario, including received signal strength (RSS), center frequency, integration of multiple signal sources, etc. The dynamic needs of the configuration are mainly changing between different scenarios, implemented tasks, and SoOPs. When the SoOP remains unchanged, its dynamic configuration is only needed once to find the optimized parameters, which can reduce the waste of computing resources while achieving system applicability. The $\mathbb{B}_1 (\text{MHz}) \in \{91.2, 93.6, 96.0, 98.4, 100.8\}$ found in previous work are used in the experiments for preliminary validation of the proposed PIPS. The PRF sensor used in our experiment is RTL-SDR RTL2832U because of its cheap and easy-to-configure characters, as shown in Fig. \ref{Fig. PRF Antenna}.

\section{Experiment and Results}
\label{Sec. Experiment and Results}
This section is organized as follows. In the experimental scenario, spectrum data are collected at 60 positions for the frequency band $\mathbb{B}_1$ that has the most impact on the positioning. Using single regressors as a baseline, three ensemble learning strategies of boosting, bagging, and stacking are compared. Three criteria are used as evaluation methods. This section focuses on the setup of experimental scenarios and the comparison and evaluation of strategies and models.

\subsection{Experimental Setup} 
\label{Sec. Experimental Setup}
\begin{figure}[!h]
\centering
\includegraphics[width=0.8\linewidth]{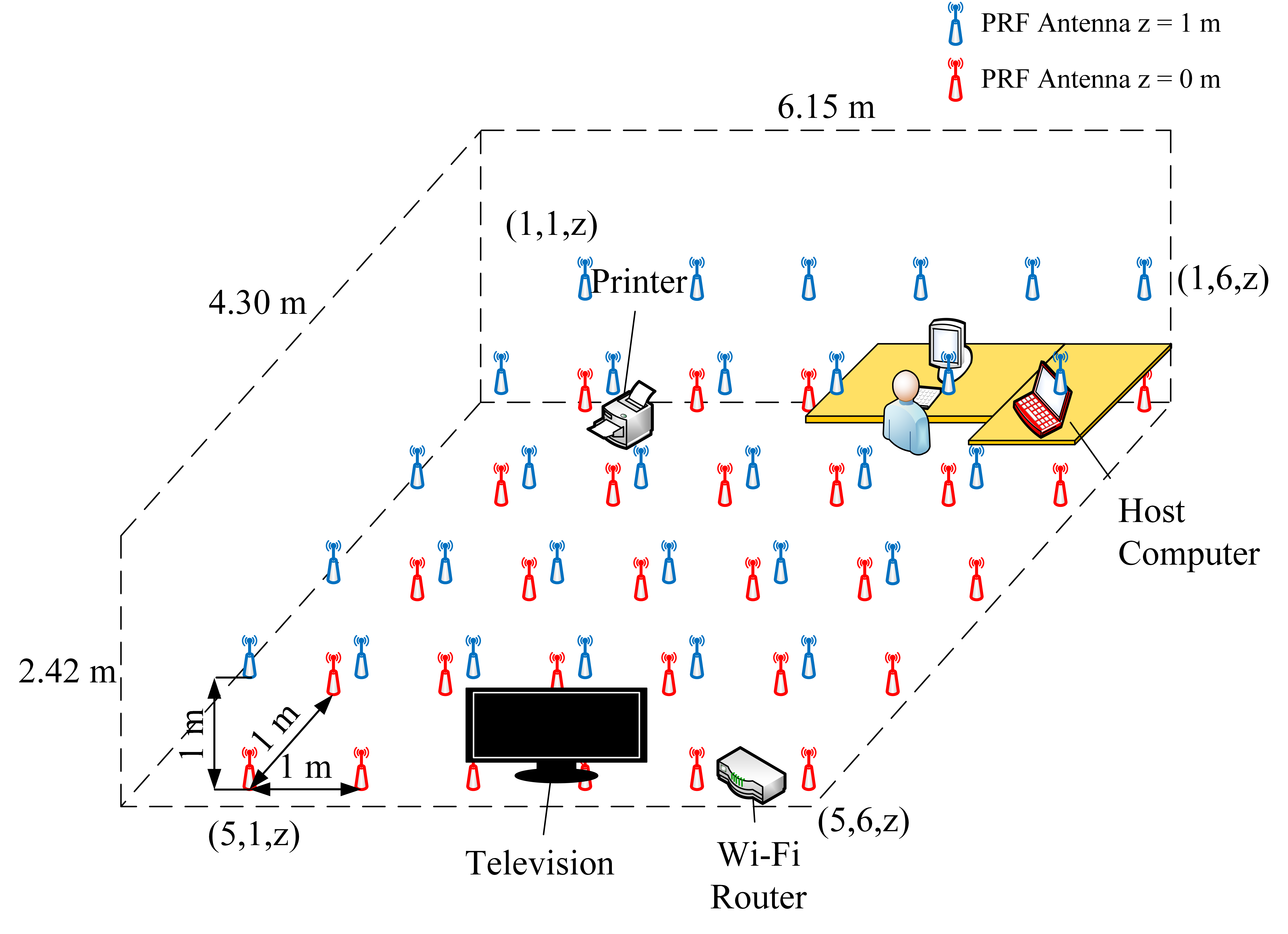}
\caption{Illustration of an indoor living room scenario is used to collect PRF data at 60 positions. The red and blue antennas are represented as 0 and 1 meters from the bottom of the antenna to the ground, respectively. Other potentially disturbing objects and human are also marked.}
\label{Fig. Experiment Scenario}
\end{figure}

Data collection is done in an indoor home scenario, as shown in Fig. \ref{Fig. Experiment Scenario}. In order to avoid the impact of sampling distance on performance and also to better compare with other state-of-the-art technologies, one meter is selected as the sampling distance in the three directions of length, width, and height. According to past experience, some sources that may have an impact on the PRF spectrum are marked in Fig. \ref{Fig. Experiment Scenario}, such as a host computer, operator, TV, Wi-Fi router, printer, etc. The experimental scenario with a length of 6.15 m, a width of 4.30 m, and a height of 2.42 m is used as a preliminary verification of the PRF positioning. We collected 100 samples at each position, and a total of 6000 samples were divided into training and test data sets in a ratio of 0.7 and 0.3. Using scikit-learn, the model was built on TensorFlow and trained with a Nvidia GeForce RTX 3080 GPU.

\subsection{Results and Evaluation}
\label{Sec. Ensemble Learning}
\begin{figure}[!h]
\centering
\includegraphics[width=0.8\linewidth]{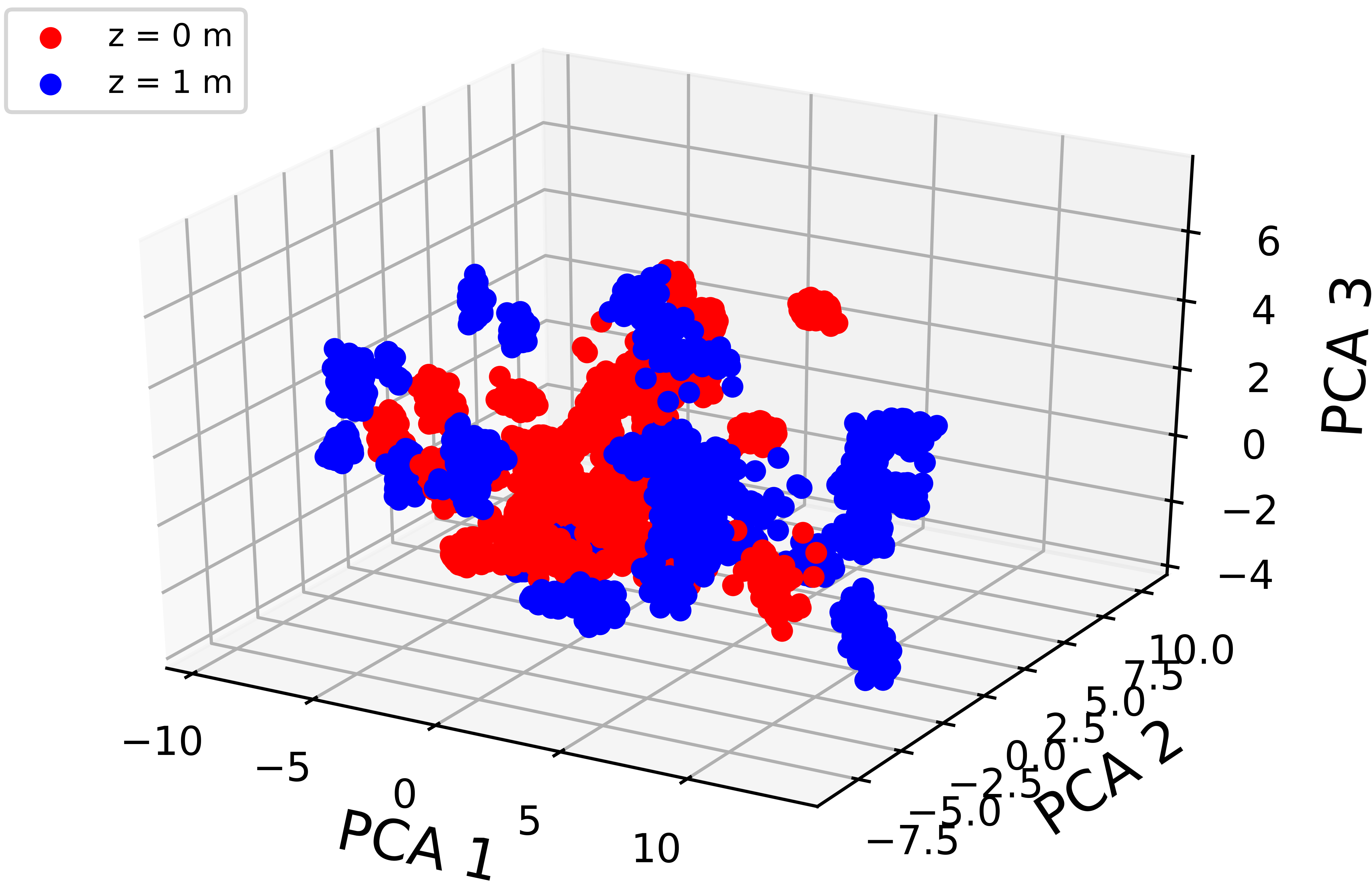}
\caption{Illustration of PRF spectrum with reduced dimensionality by PCA. The red and blue dots indicate that spectrum data was collected at 0 and 1 m from the ground, respectively.}
\label{Fig. PCA Cluster}
\end{figure}

To better demonstrate the effectiveness of the collected PRF spectrum data for positioning, principal component analysis (PCA) is used to reduce the dimensionality of the PRF spectrum data and visualization. The raw PRF spectrum is 5D since the most impacted frequency band used in data collection are five frequencies, while PCA reduces it to 3D for visualization. Using PCA is just for visualization, while the raw data set is used to train the ensemble learning model for the positioning task. Fig. \ref{Fig. PCA Cluster} shows PRF spectrum data dimensionally reduced by PCA.  Data at different positions and heights can form a cluster in the PCA space, which can prove that there are differences in data at different positions, which is also the fundamental reason for positioning.

For the proposed model, it is required to compare with the baseline in terms of performance and complexity. For ensemble learning models, some single regressors are used as the baseline, including Support Vector Regression (SVR), K Nearest Neighbors Regression (KNR), Gaussian Process Regression (GPR), Decision Trees Regression (DTR), and Multi-layer Perceptron (MLP). The performance is compared by three evaluations: Root mean square error (RMSE), coefficient of determination $R^2$, and 95$\%$ CE. RMSE is targeted at applications that require lower average errors but less stringent positioning systems, such as warehouse patrol robots. RMSE of the test data set can be expressed as
\begin{equation}
\text{RMSE} = \sqrt{\frac{\|C^\ast - \hat{C^\ast}^2\|} {n^\ast}},
\label{Eq. RMSE}
\end{equation}
where $C^\ast \in \mathbb{R}^{n^\ast \times 3}$ is the label of test data set, $\hat{C^\ast}$ is the estimated label obtained by ensemble model. 95$\%$ CE is the corresponding error when the cumulative distribution function of RMSE reaches 95$\%$, which can be expressed as
\begin{equation}
95 \% \text{CE} = F^{-1}_{\text{RMSE}}(0.95),
\label{Eq. CE}
\end{equation}
where $F$ is the cumulative distribution function of RMSE, the 95$\%$ CE is aimed at systems that are more critical to accuracy, such as firefighting robots. It requires a higher confidence level to limit the robot's error to a strict value. The time complexity is considered to be equivalent to model fitting time. Coefficient of determination and time complexity are not our main concerns. Since the proposed PIPS is an application system, it is necessary to pay more attention to the customer-oriented performance of the application. Each model was trained with its default parameters for initial comparison. The performance and complexity of some regressors are shown in Table \ref{Table 1}.
\begin{table}
\centering
\caption{Single regressors to implement positioning tasks and serve as baselines for PIPS.}\label{Table 1}
\begin{tabular}{|c|c|c|c|c|}
\hline
Regression &  RMSE (m) & $R^2$ & 95$\%$ CE (m) & Time (s)\\
\hline
SVR & 1.229 & 0.777 & 2.214 & 1.026\\
KNR & 0.268 & 0.986 & 0.412 & 0.002\\
GPR & 0.612 & 0.967 & 1.248 & 1.508\\
DTR & 0.603 & 0.930 & 1.111 & 0.016\\
MLP & 1.506 & 0.562 & 2.534 & 2.104\\
\hline
\end{tabular}
\end{table}

It can be seen from Table \ref{Table 1} that KNR has the best performance, which will be used as the baseline for PIPS to compare with the ensemble learning strategy. Different models under three ensemble learning strategies are used to train on our positioning data set. For serial boosting strategies, there are three main extensions, including Adaptive Boosting Regression (ABR), Gradient Boosting Regression (GBR), and Histogram-based GBR (HGBR). ABR makes the posterior estimator focus more on samples that cannot be solved by the prior estimator through an adaptive weight method. Both GBR and HGBR are ensembles of regression trees that use a loss function to reduce the error of the previous estimator. According to the results in Table \ref{Table 1}, we selected four models with different accuracies, namely SVR, KNR, GPR, and DTR, for further analysis. Table \ref{Table 2} shows the model performance under the boosting strategy.
\begin{table}
\centering
\caption{Performance of ensemble learning models under the boosting strategy.}\label{Table 2}
\begin{tabular}{|c|c|c|c|c|c|}
\hline
Ensemble Strategy & Base Estimator &  RMSE (m) & $R^2$ & 95$\%$ CE (m) & Time (s)\\
\hline
ABR   & SVR & 0.828 & 0.881 & 1.419 & 88.368\\
~     & KNR & 0.324 & 0.985 & 0.095 & 2.859\\
~     & GPR & 0.825 & 0.859 & 1.442 & 278.900\\
~     & DTR & 0.324 & 0.983 & 0.095 & 2.193\\
GBR   & DTR & 0.807 & 0.879 & 1.575 & 1.698\\
HGBR  & DTR & 0.457 & 0.960 & 1.027 & 1.161\\
\hline
\end{tabular}
\end{table}

It can be seen that the ensemble learning model under the boosting strategy has no advantage in RMSE compared to a single regressor, but it greatly reduces 95$\%$ CE, especially for ABR with KNR and DTR as base estimators. This means that most of the samples have errors less than 0.095, but there are also a few samples with large errors that increase the value of RMSE. Boosting strategies are effective in reducing the mode of error. For the bagging strategy, the base estimator is also a crucial parameter. In addition to the general bagging model, Random Forest Regression (RFR) and Extremely Randomized Trees (ERT) as bagging variants and extensions of DTR are also included as part of the comparison. Table \ref{Table 3} shows the performance of the models under the bagging strategy.
\begin{table}
\centering
\caption{Performance of ensemble learning models under the bagging strategy.}\label{Table 3}
\begin{tabular}{|c|c|c|c|c|c|}
\hline
Ensemble Strategy & Base Estimator &  RMSE (m) & $R^2$ & 95$\%$ CE (m) & Time (s)\\
\hline
Bagging & SVR & 1.124 & 0.775 & 2.116 & 5.028\\
~       & KNR & 0.265 & 0.989 & 0.423 & 0.372\\
~       & GPR & 0.623 & 0.928 & 1.323 & 41.264\\
RFR     & DTR & 0.418 & 0.966 & 0.964 & 0.934\\
ERT     & DTR & 0.299 & 0.966 & 0.710 & 0.304\\
\hline
\end{tabular}
\end{table}

Through the comparison of Table \ref{Table 1} and Table \ref{Table 3}, it can be found that - whether it is KNR with the best accuracy or SVR with poor accuracy, the bagging strategy cannot significantly further improve its accuracy. The final prediction of the bagging strategy will be related to each base estimator, that is, it will also be affected by the base estimator with poor accuracy. The stacking strategy aggregates base estimators through the final estimator, which gives different weights to base estimators. We use the ten previously mentioned regressors, including the ensemble learning model as the base estimator, and then test the performance of these regressors as the final estimator. The regression results under the stacking strategy are shown in Table \ref{Table 4}.
\begin{table}
\centering
\caption{Performance of ensemble learning models under the stacking strategy.}\label{Table 4}
\begin{tabular}{|c|c|c|c|c|c|}
\hline
Ensemble Strategy & Final Estimator &  RMSE (m) & $R^2$ & 95$\%$ CE (m) & Time (s)\\
\hline
Stacking & SVR & 0.271 & 0.988 & 0.463 & 97.281\\
~        & KNR & 0.259 & 0.990 & 0.446 & 92.678\\
~        & GPR & 2.115 & 0.273 & 3.924 & 97.241\\
~        & DTR & 0.327 & 0.984 & 0.086 & 93.218\\
~        & MLP & 0.263 & 0.990 & 0.459 & 95.106\\
~        & ABR & 0.334 & 0.984 & 0.258 & 97.657\\
~        & \textbf{GBR} & \textbf{0.258} & \textbf{0.990} & \textbf{0.317} & \textbf{94.338}\\
~        & HGBR & 0.254 & 0.990 & 0.371 & 95.478\\
~        & RFR & 0.255 & 0.990 & 0.431 & 93.835\\
~        & ETR & 0.259 & 0.990 & 0.334 & 93.808\\
\hline
\end{tabular}
\end{table}

The stacking strategy affords the use of any model as the base estimator, so the stacking strategy can also be a strategy that integrates ensemble learning models. The results show that the stacking strategy has an advantage in performance compared to the bagging strategy, which is because the final estimator can adaptively aggregate all the base estimators. However, the stacking strategy is not dominant compared to the boosting strategy. Although stacking is stronger than boosting in RMSE and $R^2$, the time complexity is dozens of times. 

After experiments, we found that the Stacking strategy gave the best results. Compared to the baseline, the proposed ensemble learning strategy has considerable improvement on 95$\%$ CE. In particular, the stacking strategy with DTR as the final estimator can reduce the 95$\%$ CE by 92.3$\%$. Although 95$\%$ of the samples have relatively low errors, the average RMSE is still high, which means that minority samples with a proportion of 5$\%$ or less have considerable errors. These samples may have received interference, such as the movement of the human body, the effect of metal or liquid shielding on the PRF spectrum, etc. Therefore, GBR as the final estimator is considered as the global optimal solution, which outperforms the baseline in all aspects.

\section{Discussion}
\label{Sec. Discussion}
DDDAS is crucial for PIPS, the main purpose of the DDDAS framework is to find the optimal solution for the positioning task in the target scenario. We implement pre-sampling in the target scenario and then use SHAP to analyze the collected samples and find the optimal frequency band, step size, and sampling rate. PIPS under the DDDAS framework has three advantages. Firstly and most importantly, the sampling time is reduced by 98$\%$. We reduced the 400 frequencies to 5 frequencies under the DDDAS framework. If we do not use the DDDAS framework to find the optimal frequency, data collection over the full frequency band will waste a huge amount of time. Secondly, the redeployment time of the sensor is also greatly reduced. The proposed PIPS system also has excellent redeployment capabilities in new scenarios, thanks to the DDDAS framework on frequency band $\mathbb{B}$, step size $\Delta$, sampling rate $R_s$ optimization. To achieve the accuracy and sampling resolution described above, the time resource required for redeployment is around 300 $s/m^3$.The training time is negligible compared to the PRF data sampling time. 

Thirdly, it can potentially improve accuracy and reliability. The PIPS system uses RSS in the five most sensitive frequencies, especially since this passive RF technology can capture signatures from scenarios such as metal parts in house structures or liquids. So basically, the PRF signal collected in each scenario is unique. On the one hand, there are inevitably some interferences in the full frequency band, including natural noise and artificial signals. These noise signals are random and abrupt, which is not conducive to the stability of a positioning system. On the other hand, we don't want to include any unnecessary features in the samples. In this task, we did not use deep learning but just traditional machine learning. Traditional machine learning cannot adaptively assign weights, so unnecessary and cluttered features obviously affect the accuracy of classification. Therefore, collecting data in the most sensitive frequency bands for positioning can effectively avoid these possible interferences and reduce feature complexity to improve accuracy and reliability.

\section{Conclusion}
\label{Sec. Conclusion}
This paper proposes a PIPS under the DDDAS framework to solve the 3D positioning problem. Three ensemble learning strategies and their various variants and extensions are used to train on the collected data set. The experimental results show that the proposed ensemble learning strategy has an RMSE of 0.258 meters, an $R^2$ of 0.990, and a 95$\%$ CE of 0.317 meters, which is much better than the baselines. PIPS under the DDDAS framework is considered a potential application in specific scenarios, such as robot-patrolled factories or warehouses, due to its efficient redeployment and high accuracy.

For future work, dimensionality reduction is a potential research direction. The current work is limited to the frequency selection technology. We selected the most sensitive frequency band from the 400 frequencies in the full frequency band under the DDDAS framework. However, dimensionality reduction, while similar in terms of results, has different effects. For the dimensionality reduction method, although it can reduce the complexity of the data, it cannot reduce the sampling time. The benefits of PCA lie in privacy considerations and visualization applications. In internet of things (IoT) applications, performing PCA processing locally can reduce the dimension of data so that customer privacy can be protected after uploading to the cloud. PCA is able to reduce multi-dimensional data to three-dimensional or two-dimensional to enable visualization applications.

\subsubsection{Acknowledgements} Thanks to Dr. Erik Blasch for concept development and co-authoring the paper. This research is partially supported by the AFOSR grant FA9550-21-1-0224. The views and conclusions contained herein are those of the authors and should not be interpreted as necessarily representing the official policies or endorsements, either expressed or implied, of the Air Force Research Laboratory or the U.S. Government.

\end{document}